\documentclass[10pt,conference]{IEEEtran}
\IEEEoverridecommandlockouts
\usepackage[T1]{fontenc}
\usepackage{cite}
\usepackage{amsmath,amssymb}
\usepackage{graphicx}
\usepackage{booktabs}
\usepackage{array}
\usepackage{enumitem}
\usepackage{csquotes}
\usepackage{tabularx}
\usepackage{xcolor}
\usepackage{colortbl}
\usepackage{xurl}
\usepackage[hidelinks]{hyperref}
\usepackage{microtype}
\usepackage{pifont}
\usepackage{soul} %
\definecolor{groupshade}{gray}{0.93}
\newcolumntype{Y}{>{\raggedright\arraybackslash}X}
\newcommand{\tact}{\texorpdfstring{\textsc{tact}}{TACT}}
\newcommand{\lite}{\textsc{tact}-Lite}
\newcommand{\full}{\textsc{tact}-Full}

\newcommand{\fullonly}[1]{\textit{\full\ only: #1}}

\usepackage[normalem]{ulem}

\newcommand{\dmCorpusThreads}{256}
\newcommand{\dmZhangThreads}{303}
\newcommand{\dmDeletedThreads}{47}
\newcommand{\dmDevThreads}{64}
\newcommand{\dmEvalThreads}{192}
\newcommand{\dmEvalLabels}{183}
\newcommand{\dmEvalLabelledThreads}{128}
\newcommand{\dmZhangCategories}{48}
\newcommand{\dmCodesLiteCandidate}{80}
\newcommand{\dmCodesFullCandidate}{87}
\newcommand{\dmCodesReviewed}{72}
\newcommand{\dmCBLiteRecall}{0.82}
\newcommand{\dmCBLitePrecision}{0.68}
\newcommand{\dmCBLiteFone}{0.75}
\newcommand{\dmCBLiteReached}{151}
\newcommand{\dmCBFullRecall}{0.84}
\newcommand{\dmCBFullPrecision}{0.71}
\newcommand{\dmCBFullFone}{0.77}
\newcommand{\dmCBFullReached}{154}

\newcommand{\dmMapLiteCatsPerCode}{1.7}

\newcommand{\dmMapFullCatsPerCode}{1.5}

\newcommand{\dmMapRevUnmapped}{9}

\newcommand{\dmCodeLiteAlphaMedian}{0.77}

\newcommand{\dmCodeFullAlphaMedian}{0.87}

\newcommand{\dmCodeLiteKappa}{0.78}
\newcommand{\dmCodeLiteKappaClGp}{0.78}
\newcommand{\dmCodeLiteKappaClGe}{0.82}
\newcommand{\dmCodeLiteKappaGpGe}{0.79}

\newcommand{\dmCodeFullKappa}{0.86}
\newcommand{\dmCodeFullKappaClGp}{0.89}
\newcommand{\dmCodeFullKappaClGe}{0.89}
\newcommand{\dmCodeFullKappaGpGe}{0.88}

\newcommand{\dmFCLabels}{182}
\newcommand{\dmFCLiteRecall}{0.70}
\newcommand{\dmFCLitePrecision}{0.72}
\newcommand{\dmFCLiteFone}{0.71}
\newcommand{\dmFCLiteReached}{127}

\newcommand{\dmFCFullRecall}{0.71}
\newcommand{\dmFCFullPrecision}{0.72}
\newcommand{\dmFCFullFone}{0.72}
\newcommand{\dmFCFullReached}{130}

\newcommand{\dmCodeFullIssue}{8}

\newcommand{\dmProvQuotes}{1405}

\newcommand{\dmCostLiteTotal}{35.26}

\newcommand{\dmCostFullTotal}{86.00}

\newcommand{\dmCallsLiteDisc}{19}

\newcommand{\dmCostLitePerHundred}{13.77}
\newcommand{\dmCallsFullDisc}{31}

\newcommand{\dmCostFullPerHundred}{33.59}
\newcommand{\dmCostJevScreen}{0.24}

\newcommand{\dmScreenShareShown}{99}

\newcommand{\dmBaseRecall}{0.62}
\newcommand{\dmBasePrecision}{0.68}
\newcommand{\dmBaseFone}{0.65}

\newcommand{\dmBaseMatched}{113}

\newcommand{\dmBaseBootLite}{97}
\newcommand{\dmBaseBootFull}{99}

\begin{document}

\title{TACT: A Pragmatic Approach to Qualitative Analysis with an AI Council}

\author{\IEEEauthorblockN{Rudrajit Choudhuri\IEEEauthorrefmark{1},
Christian Bird\IEEEauthorrefmark{2},
Carmen Badea\IEEEauthorrefmark{2},
Anita Sarma\IEEEauthorrefmark{1}}
\IEEEauthorblockA{\IEEEauthorrefmark{1}Oregon State University, Corvallis, OR, USA, \{choudhru, anita.sarma\}@oregonstate.edu}
\IEEEauthorblockA{\IEEEauthorrefmark{2}Microsoft Research, Redmond, WA, USA, \{cbird, cabadea\}@microsoft.com}}

\IEEEaftertitletext{\vspace{-1.1\baselineskip}}
\maketitle

\begin{abstract}
Qualitative data can tell software teams what their stakeholders need, yet rigorously analyzing thousands of qualitative responses takes expert time that few teams can afford or justify. Teams therefore often face a false choice---leave the responses unanalyzed, or hand them to a large language model and accept its interpretation and judgment. In this paper, we take a pragmatic approach to AI-assisted qualitative analysis and present \tact{} (Traceable AI Council for Thematic analysis), a procedure that enables such analysis at scale while keeping teams in control of interpretation and judgment. \tact{} comes in two variants: \lite{}, the core AI council procedure, and \full{}, which adds adversaries (models from other model families that critique each model's output) and a decision model that surfaces high-uncertainty code assignments. We evaluated \tact{} on an independent dataset of \dmCorpusThreads{} Stack Overflow threads qualitatively coded by other researchers. Against their labels, \tact{}'s candidate codebook achieved an F1 of \dmCBLiteFone{}--\dmCBFullFone{} and its final coding an F1 of \dmFCLiteFone{}--\dmFCFullFone{}. On a held-out sample, agreement between our own manual coding and \tact{}'s coding was Cohen's $\kappa$ of 0.79--0.85. \tact{} is also economical: model calls cost \$\dmCostLitePerHundred{} (\lite{}) to \$\dmCostFullPerHundred{} (\full{}) per 100 threads. Finally, drawing on our experience building and validating \tact{}, we identify pitfalls in AI-assisted qualitative analysis, describe safeguards to address them, and offer guidance for teams and researchers applying \tact{} in practice.

\end{abstract}


\section{Introduction}

\begin{quote}\small\itshape
``You usually can't change people's minds by the intellect. You've got to find something that reaches into their hearts.''\par\hfill\upshape Jane Goodall~\cite{bell2010goodall}
\end{quote}

Metrics, telemetry and survey ratings can show a software team what is happening, but the reasons usually sit in the accounts of the people behind those numbers. Whether a team is running user research or an empirical study, hearing people describe where their tools and practices fail them, or what they wish they had, can help the team understand the people it builds for and what to change. For example, Nahar et al.\ interviewed practitioners building LLM-enabled products and identified engineering challenges and emerging practices that informed a broader survey~\cite{nahar2025beyond}. Their analysis then led to an internal Microsoft workshop on these challenges. In our experience, accounts like these have the potential to persuade management and influence its priorities.

However, producing such understanding takes sustained analytic work. Researchers analyze large quantities of unstructured text, find patterns, develop categories, revisit the material and the categories, compare interpretations, and examine accounts that contradict an emerging explanation. This work requires time, methodological expertise and coordination among researchers~\cite{gao2024collabcoder}, and its cost grows with every response collected. Among practitioners, limited time or resources can leave valuable material only partly analyzed, and even a team that wants rigor often finds its cost hard to justify.

As a result, a team facing hundreds of qualitative responses, with little time and few resources to analyze them, rarely attempts a careful analysis. In practice, it often leaves the responses unanalyzed or hands them to a large language model (LLM). Unfortunately, such an analysis offers little way to check its rigor or for humans to retain interpretive judgment. Further, the model can return themes that are not grounded in respondents' actual words, and it is opaque about how each response was coded. Moreover, access to a model does not necessarily supply the oversight that careful analysis requires.

We argue that forgoing qualitative analysis or handing it entirely to an LLM are not the only two choices. In this paper, we take a pragmatic approach to using AI responsibly in qualitative analysis. We present \tact{} (Traceable AI Council for Thematic analysis), which operationalizes bounded delegation~\cite{choudhuri2026copilot} to a council of AI models for qualitative analysis in software engineering (SE) research and practice. We distilled it from three of our prior studies~\cite{choudhuri2026copilot,choudhuri2026youshallnotpass,choudhuri2026coworker}. The delegation is bounded in that models carry out only explicit pattern-discovery and coding tasks, each grounded in evidence that researchers can check, while researchers retain every interpretive decision. 

\begin{figure*}[t]
\centering
\includegraphics[width=0.9\textwidth]{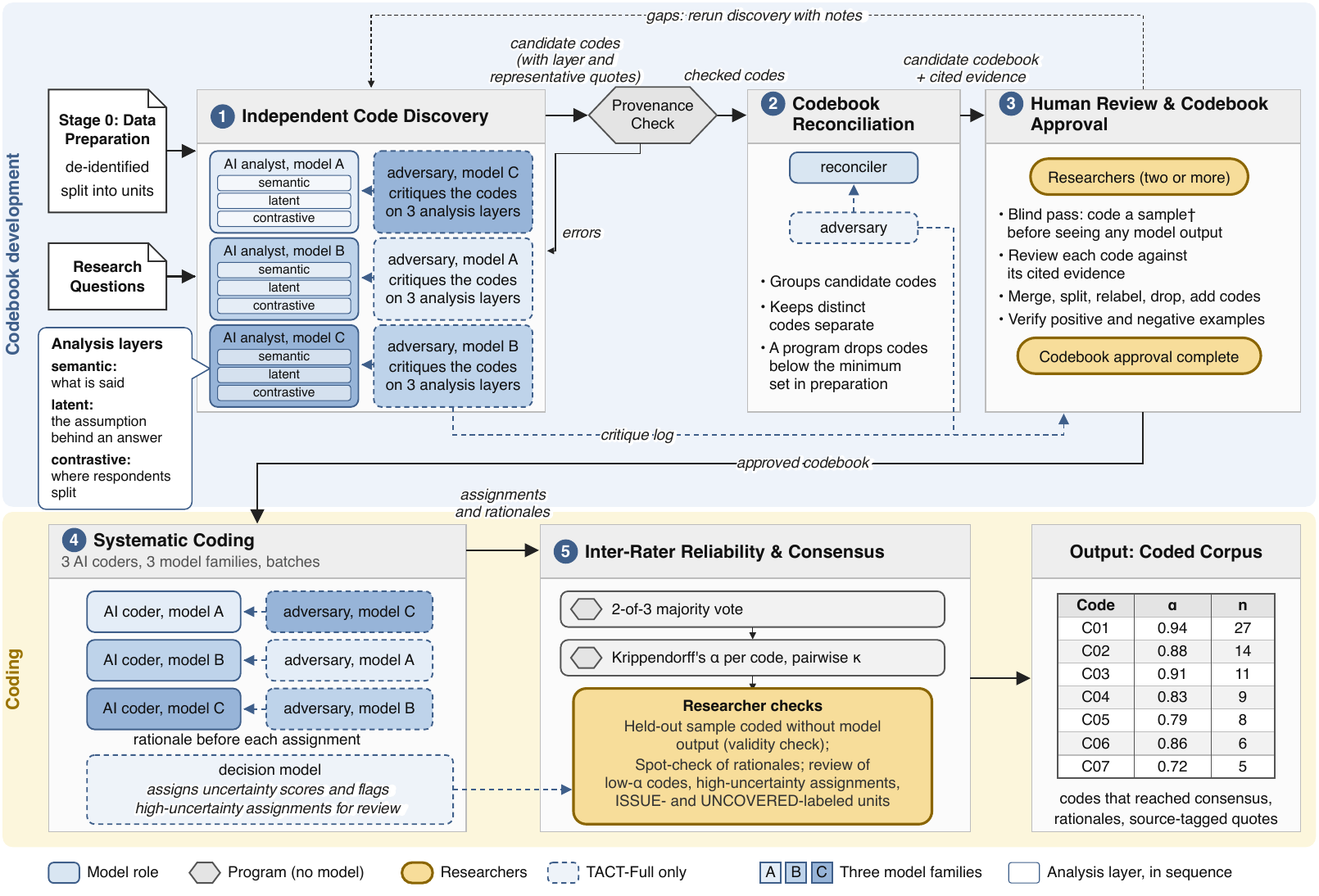}
\caption{\small The \tact{} procedure. After data preparation (Stage~0), Stages~1--3 develop the codebook: AI analysts identify candidate codes, a reconciler merges them, and researchers approve the codebook by coding a blind-pass sample themselves (\dag). Stages~4--5 apply the codebook to the corpus: AI coders code every unit, a program computes consensus and reliability, and researchers check the result against their own coding of a held-out sample. Models A, B, and C come from three different model families. Solid elements are \lite{}; dashed elements are \full{}'s additions, in which adversaries from another model family critique each AI analyst, reconciler, and each AI coder, and a decision model assigns uncertainty scores to code assignments and flags high-uncertainty ones for review.}
\label{fig:council}
\end{figure*}

Figure~\ref{fig:council} shows the procedure: after data preparation, it runs in five stages: (1)~three \emph{AI analysts}, each from a different model family, independently discover candidate codes, each with the evidence from the data that supports it, and a deterministic program verifies that every piece of quoted evidence appears verbatim in the data; (2)~a \emph{reconciler} model then merges these candidate codes into a candidate codebook; (3)~researchers, who first code a sample of the responses themselves (10\% by default) before seeing any AI output, review the candidate codes against their evidence, refine them and approve the codebook; (4)~three \emph{AI coders}, again one from each model family, code the corpus against the approved codebook and write a rationale for each code they assign; and (5)~finally, a deterministic program settles each response's code assignments by majority vote (consensus) and measures the AI coders' agreement (inter-rater reliability), and researchers compare the result with their own coding of a held-out sample. \tact{} keeps provenance from end to end, so each code can be traced to the responses behind it and each decision to the model or researcher who made it.

\tact{} can thus make rigorous qualitative analysis accessible to teams with some qualitative training but limited time and resources. It comes in two variants: \lite{}, the core procedure, and \full{}. \full{} adds adversaries from other model families, which critique the output of each AI analyst, the reconciler, and each AI coder. It also adds a decision model that surfaces uncertainty in code assignments.

We show \tact{}'s reliability and economic viability on an independent dataset. In 2023, Zhang et al.\ collected Stack Overflow threads about GitHub Copilot and coded them manually~\cite{zhang2023demystifying}. We ran both \tact{} variants on \dmCorpusThreads{} of these threads and compared \tact{}'s codes with Zhang et al.'s labels. \tact{}'s candidate codebook agreed with their labels at an F1 of \dmCBLiteFone{}--\dmCBFullFone{}, and its final coding at an F1 of \dmFCLiteFone{}--\dmFCFullFone{}. The three AI coders agreed with one another (median Krippendorff's $\alpha$ of \dmCodeLiteAlphaMedian{}--\dmCodeFullAlphaMedian{}), and our own coding of a held-out sample agreed with \tact{}'s coding at Cohen's $\kappa$ of 0.79--0.85. \tact{} is also economical: model calls cost \$\dmCostLitePerHundred{} (\lite{}) to \$\dmCostFullPerHundred{} (\full{}) per 100 threads in model calls. We release \tact{} so that teams can run it on their own data where their inquiry suits AI assistance~\cite{ernst2026genai}.

Overall, this paper makes three contributions for SE practitioners and researchers undertaking qualitative analysis:
\begin{itemize}[leftmargin=*,topsep=2pt,itemsep=1pt]
\item \textbf{\tact{}, an AI council procedure for qualitative analysis in SE research.} We describe what the models do at each stage, which checks programs run, and where researchers decide, in two variants, \lite{} and \full{} (Section~\ref{sec:procedure}).
\item \textbf{Practical guidance grounded in experience.} Drawing on our experience of building and using \tact{} in three studies, and on its validation on an independent dataset, we list common pitfalls in AI-assisted qualitative analysis and the safeguards against them (Sections~\ref{sec:procedure}--\ref{sec:discussion}).
\item \textbf{Replication artifacts.} We release a replication package with the prompts, scripts, and the validation corpus~\cite{tactsupplement}, so that readers can check our validation and teams can run \tact{} on their own data.
\end{itemize}

\section{Related Work}
\label{sec:background}

Thematic analysis is a widely used method for identifying, analyzing and reporting patterns of meaning, or themes, in qualitative data~\cite{braun2006using}. Its rigor rests on codes that are grounded in the data and on researchers' interpretation of them. Because researchers analyze and interpret every response, the work grows with the data. The qualitative research community has long built tools that help researchers organize and share their coding~\cite{lewins2007software,marathe2018semi,gao2024collabcoder}. More recently, researchers have begun to test whether LLMs can perform parts of this work~\cite{xiao2023supporting,chew2023llmassisted,dai2023llminloop,depaoli2024performing,bano2024large,barros2025large,marston2026can,rush2026evaluating}.

Early studies found that a single LLM can code deductively, against a codebook, with useful accuracy~\cite{xiao2023supporting}, and more accurately when asked to provide its reasoning before assigning a code~\cite{dunivin2024scalable}. Models have also been shown to be more dependable on descriptive codes, which capture what respondents say, than on interpretive codes, which capture what they mean~\cite{morgan2023exploring,li2026deductive}. However, a single model can also bias an analysis. For example, Ashwin et al.\ found that a model made more coding errors for some groups of respondents than for others~\cite{ashwin2026using}, and Gao et al.\ found that one model shared by several human coders homogenized their codes~\cite{gao2024coaicoder}. Furthermore, Huang et al.\ found that without external feedback, such as a second model or another perspective, a model often cannot correct its own reasoning~\cite{huang2024cannot}. 

In response, research has started combining the judgments of several models. For example, Verga et al.\ found that a panel of models from different model families judged more accurately than a single judge, at lower cost~\cite{verga2024replacing}, and Wu et al.\ found that a heterogeneous council of models reduced hallucination~\cite{wu2026council}. Similarly, Dunivin et al.\ found that a second model re-examining the text and the first model's rationale cut false positives in qualitative coding by a wide margin~\cite{dunivin2026selfreflection}. Multi-agent systems for thematic analysis bring these ideas together and assign roles to several agents, such as coding agents with an aggregator~\cite{qiao2025thematiclm}, a coding agent with peer-debriefing agents~\cite{lin2026agent}, and agents that code, debate and reconcile~\cite{kim2026how}. Still, models can share errors. Kohli found that nine leading model judges erred so similarly that they gave about as much information as two independent judges~\cite{kohli2026nine}, so agreement among models shows that they applied a codebook consistently but leaves open whether the codes are right~\cite{liu2026agreement}. Finally, models can also sway the people who check them, as Schroeder et al.\ found that model suggestions anchor the people who see them~\cite{schroeder2025just}. 

Building on this work, we add responsible safeguards to the \tact{} procedure. First, to limit the bias of any single model, AI models from three model families code the data independently, and each provides a rationale before assigning a code. Second, to give each model the external feedback it cannot give itself, \full{} adds adversaries from other model families that critique each model's output. Third, because agreement among models cannot show that the codes are right, researchers verify that each code the council discovers is substantiated by its cited evidence, and compare the council's coding with their own coding of a held-out sample. Finally, to avoid anchoring, human analysts code a sample of the data before they see any AI output, and researchers remain responsible for all interpretive decisions throughout.

Overall, while research has examined whether AI can perform parts of qualitative analysis, SE researchers still lack a concrete, end-to-end, pragmatic procedure that they can apply to their own data reliably and economically. We aim to fill this gap with \tact{}, which we refined in three of our studies and validated on an independent dataset, as detailed next.

\vspace{-1mm}
\section{Author Positionality}
\label{sec:positionality}

Our research team consists of SE and HCI researchers with an average of 10 years of experience conducting mixed-methods empirical studies in academia and industry. Based on this experience, we take a pragmatic stance toward AI-assisted qualitative analysis, in which we judge a method by whether it helps solve the problem at hand~\cite{easterbrook2008selecting}. Within this pragmatic stance, we adopt an interpretivist view of meaning, in which meaning is constructed through participants' accounts as well as researchers' experience, and we treat both as valuable sources of understanding~\cite{charmaz2014constructing,braun2022conceptual}. These commitments shaped \tact{}. In our procedure, AI models carry out explicit analytic tasks and provide evidence for their output, while researchers retain final authority over interpretation. We recognize that researchers hold different epistemological stances regarding the use of AI in qualitative analysis, and we respect those views. Our claim in this paper is that \tact{} offers teams that already use AI in qualitative analysis an alternative for conducting such analysis more responsibly.

\vspace{-1mm}
\section{TACT, a Bounded AI Council Procedure}\label{sec:procedure}

Our goal was to build a procedure that lets researchers analyze qualitative data responsibly at scale while retaining final authority over interpretation. To do so, we developed \tact{}, distilling it from our experience in three of our prior studies. In \tact{}, \textbf{AI models} discover and apply patterns in the data, \textbf{deterministic programs} (scripts) check the provenance of cited evidence, and \textbf{researchers} retain final authority over interpretation. \tact{} uses an AI \emph{council} of models from three frontier model families, Anthropic's Claude, Google's Gemini and OpenAI's GPT (in our validation, Claude Opus~5, Gemini 3.6 Flash and GPT-5.6), so that no single model family's blind spots or biases dominate the analysis~\cite{wu2026council}. Each model receives only the inputs it needs (Table~\ref{tab:roles} lists each role and its inputs), and models performing the same task do so independently, without seeing each other's output.

Figure~\ref{fig:council} shows the procedure, which comes in two variants. \lite{} is the core procedure, which we used in three earlier studies (Table~\ref{tab:studies}), first in~\cite{choudhuri2026copilot}. After data preparation (Stage~0), Stages~1--3 develop the codebook: AI analysts discover candidate codes with evidence (Stage~1), a reconciler merges them into a candidate codebook (Stage~2), and researchers review and approve it after first coding a sample themselves (Stage~3). Stages~4--5 apply the approved codebook to the corpus: AI coders code every unit (Stage~4), and a program combines their decisions and measures their agreement (inter-rater reliability), while researchers spot-check the codes against their rationales and compare the results with their own coding of a held-out sample (Stage~5). \full{} adds two components, drawn as dashed boxes in Figure~\ref{fig:council}. The first is \emph{adversaries}, models from another model family that critique each model's decisions (see Stages~1, 2 and~4). The second is a \emph{decision model}, which assigns an uncertainty score to each code assignment, as described in Stage~4. We now describe each stage in turn.

\begin{table}[t]
\footnotesize
\caption{Roles and steps in \tact{} and their inputs. AI analysts and AI coders: one per model family. Adversaries: another model family. $^{P}$\,A program. $^{R}$\,Researchers. $^{F}$\,\full{} only.}
\label{tab:roles}
\setlength{\tabcolsep}{3pt}
\begin{tabularx}{\columnwidth}{@{}>{\raggedright\arraybackslash}p{3.9cm}Y@{}}
\toprule
Role or step ($n$) & Inputs \\
\midrule
\rowcolor{gray!18}AI analyst~(3) & the corpus and the common prompt; no other AI analyst's output \\
Discovery adversary$^{F}$~(3) & the corpus, one AI analyst's output \\
\rowcolor{gray!18}Provenance check$^{P}$ & the AI analysts' output and the corpus \\
Reconciler~(1) & candidate codes that passed the provenance check (in \full{}, with their critique logs); the overlap table \\
\rowcolor{gray!18}Recon.\ adversary$^{F}$~(1) & the merge plan, candidate codes and overlap table \\
Human review$^{R}$~($\ge2$) & all stage outputs and critique logs, after the blind pass \\
\rowcolor{gray!18}AI coder~(3) & the approved codebook and a batch with context; no other AI coder's output \\
Coding adversary$^{F}$~(3) & one AI coder's batch and assignments, the codebook \\
\rowcolor{gray!18}Decision model$^{F}$~(1) & one unit, every code \\
Consensus and reliability$^{P}$ & all AI coders' final assignments \\
\bottomrule
\end{tabularx}
\vspace{-2mm}
\end{table}

\textbf{\textit{Stage 0: Data preparation.}} Researchers first decide how the data may be used and how the coding will be assessed. Specifically, they decide which model providers may receive the data and how much it must be de-identified under company policy and respondents' consent. They then fix the analysis thresholds, i.e., the minimum number of distinct respondents whose responses must support a code (three by default) and the minimum agreement among the AI coders on each code, measured with Krippendorff's $\alpha$~\cite{krippendorff2018content}. We consider the default of 0.7 a baseline suited to exploratory studies~\cite{lombard2002content,stemler2004comparison}. Researchers can raise it to fit the stakes of their study, e.g., to 0.8 for conservative conclusions~\cite{krippendorff2018content}. A code whose $\alpha$ falls below the threshold returns to the researchers, who either refine its definition and have the AI coders recode it, drop the code, or keep it with a stated reason, and report the decision. For \full{}, researchers also fix the decision model's thresholds, detailed in Stage~4. Finally, they state each research question and what they want to learn from it in the \emph{prompts} that instruct the models.

A deterministic program then prepares the corpus. It splits the corpus into \emph{units} of analysis, e.g., a response to one survey question, an interview excerpt, or a discussion thread. Before any model provider receives the data, the program de-identifies it and gives each unit a unique identifier, so that evidence can be traced across stages. It attaches each respondent's other answers as \emph{context}, which models see but do not code. Finally, it draws two separate random samples, stratified by question, that researchers code without seeing any model output. The first is a \emph{blind-pass sample} of 10\% of the units (at least 50, at most 150), from which researchers develop their own codes to check the candidate codebook against (Stage~3). The second is a \emph{held-out sample} of at least 10\% of the units, against which they check the council's coding (Stage~5). %

\textbf{\textit{Stage 1: Independent code discovery.}} Three AI analysts, one from each model family, independently analyze the whole corpus to discover candidate themes and codes. Each AI analyst analyzes the corpus at three \emph{analysis layers} per research question: \emph{semantic} (what respondents say at face value, in their own terms), \emph{latent} (the assumptions or ways of working behind answers, so that answers worded differently can share a code~\cite{braun2006using}), and \emph{contrastive} (where respondents' answers diverge from their overall stance, e.g., conditions they set before they would act or positions on which they disagree).

For each code, the AI analyst gives a short label, a definition in respondents' own terms, inclusion and exclusion rules for when the code applies, and positive and negative examples. It then cites the units that support the code, with representative quotes from at least three of them, so that researchers can review the code against its evidence at human review (Stage~3). If no code fits a unit, the AI analyst labels the unit as having no code instead of dropping it silently. When the corpus is too long for the model's context window, the AI analyst analyzes it in batches. The AI analysts' codes then pass through the \emph{provenance check}, a deterministic program that checks each AI analyst's cited evidence against the corpus before the reconciler sees any code. It confirms that every cited unit and quote exists in the corpus and that no unit was skipped, which catches unsubstantiated evidence that a model invented, and it returns any failure to the AI analyst for correction.

\fullonly{discovery adversaries.} To catch plausible but unsubstantiated evidence and missed units, \full{} pairs each AI analyst with an adversary that critiques its discoveries and decisions before the provenance check~\cite{dunivin2026selfreflection}. The adversary receives the corpus and the AI analyst's codes and critiques the codes at each analysis layer, pointing to the units and quotes that each critique concerns. The AI analyst accepts, partly accepts or rejects each critique with a reason and revises its codes, and a \emph{critique log} records the critiques and the AI analyst's responses. A code that the adversary found but the AI analyst did not adopt stays marked as the adversary's. The reconciler then decides on it (Stage~2) and, if it cannot resolve it, surfaces it for human review. The revised codes then pass to the provenance check, and the original codes and critique logs go on to the reconciler and then to the researchers.

\textbf{\textit{Stage 2: Codebook reconciliation.}} A reconciler combines the three AI analysts' themes and codes into one candidate codebook, merging codes that overlap, combining similar ones and refining their definitions. One reconciler suffices because this is a consolidation rather than a discovery task. In this stage, a deterministic program first builds an \emph{overlap table}, which lists, for each pair of codes from different AI analysts, the units both cite and two measures of overlap, Jaccard similarity~\cite{jaccard1912distribution} and the overlap coefficient~\cite{simpson1943mammals}. The reconciler reads this table alongside each code's definition and quotes and proposes a \emph{merge plan} that sorts the codes into groups, each of which becomes one code of the candidate codebook. It groups codes only when they share units and describe the same thing within one research question. Because an AI analyst may leave a fitting unit uncited, the reconciler weighs the definitions and quotes as well as the overlap scores and gives a rationale for each group, so that low overlap alone does not rule out a merge. A program then drops codes supported by fewer respondents than the minimum (Stage~0) and lists them. Overall, this stage passes the candidate codebook to the researchers together with the merge plan, which records how each code was formed, and the list of dropped codes, from which they can restore weakly substantiated codes if they consider them important.

\fullonly{reconciliation adversary.} Before the candidate codebook goes to human review, an adversary critiques the merge plan. It checks whether the plan merged codes that should stay apart, such as opposite positions or codes that rest on few shared units, left apart codes that should be merged, or dropped codes that have support. The reconciler answers each critique as in Stage~1 and revises the plan where needed, and the researchers also receive the original plan and the critique log for their review.

\textbf{\textit{Stage 3: Human review and codebook approval.}} In this stage, researchers review what the models produced: the candidate codebook, the merge plan and the list of dropped codes, and in \full{} also the adversaries' critique logs. Because model output cannot show what is missing and can anchor the judgment of those who review it~\cite{schroeder2025just}, researchers first open-code the blind-pass sample (Stage~0) without seeing any model output. They then compare the codes they found with the candidate codebook, noting codes the models missed and codes the models found that they did not. The blind pass can also serve as a way to immerse the researchers in the data and the participants' experiences, which may leave them better placed to catch the models' errors~\cite{bainbridge1983ironies}.

In this stage, the researchers also check whether each code's label captures what respondents describe, whether its cited evidence supports it, and whether its definition separates it from similar codes. As necessary, they split, merge, relabel, move, drop, restore or add codes, settling their differences through negotiated agreement when more than one researcher is involved~\cite{creswell2016qualitative}, and record each decision and its reason in a \emph{review log}. If no candidate code fills a gap they found, they rerun discovery (Stage~1) with notes on the missing codes. For each code, they also verify and refine the positive and negative examples that the AI analysts proposed, which give the AI coders context in Stage~4. Finally, the researchers approve the finalized codebook. Note that researchers who already have a finalized codebook can go directly from data preparation (Stage~0) to Stage~4, as we did in~\cite{choudhuri2026youshallnotpass}.

\textbf{\textit{Stage 4: Systematic coding.}} Once the codebook is approved, three AI coders, one per model family, independently code every unit of the corpus; for a long corpus, we recommend coding in batches of 20 to 50 units (25 in our implementation). Before it assigns a code to a unit, each AI coder provides a rationale that explains why the unit receives that code, citing the relevant portion of the unit, as providing a rationale first has been shown to improve model reasoning~\cite{wei2022chain}. This also lets researchers check each assignment against its evidence and rationale. AI coders may assign multiple codes to one unit but cannot add codes that are not in the codebook, which prevents code drift. When a unit fits no code, the AI coder codes it \textsc{uncovered} with its rationale; when it has a data problem, e.g., an answer to a different question, no real answer or a vague response, the AI coder codes it \textsc{issue}. Both labels are surfaced to the researchers in Stage~5, so that they can review potential gaps in the codebook and problems in the data.

\fullonly{coding adversaries and high-uncertainty assignments.} \full{} adds two steps to coding. First, as in Stage~1, an adversary critiques each AI coder's assignments, and the AI coder answers each critique and revises its assignments. Second, \tact{} uses a decision model, i.e., TypeSafe's Jev~\cite{typesafe2026systemone}, a model trained to return calibrated probabilities, to answer a yes/no question with a score between 0 and 1: how likely each code is to apply to each unit, judged from the code's definition and its inclusion and exclusion rules. These scores serve two purposes. Before coding, the decision model \emph{screens} the codes for each unit, so that each AI coder sees only the codes likely to apply to it (by default, codes scoring at least 0.10). After coding, consensus assignments that the decision model scored below 0.25, i.e., less than a one-in-four probability that the code applies, are flagged as \emph{high-uncertainty assignments} for the researchers to review.

\begin{table*}[t]
\scriptsize
\caption{\lite{} in the field: three studies that ran all or part of \lite{} before this paper. Stages: 1 discovery, 2 reconciliation, 3 human review and approval, 4 systematic coding, 5 reliability and consensus.}
\label{tab:studies}
\setlength{\tabcolsep}{3.5pt}
\begin{tabularx}{\textwidth}{@{}>{\raggedright\arraybackslash}p{1.8cm}>{\raggedright\arraybackslash}p{2.6cm}>{\raggedright\arraybackslash}p{2.3cm}>{\raggedright\arraybackslash}p{1.3cm}>{\raggedright\arraybackslash}p{2.2cm}>{\raggedright\arraybackslash}p{2.9cm}>{\raggedright\arraybackslash}X@{}}
\toprule
Study & Used for & Corpus & Stages & Models & Reliability & Human check \\
\midrule
To Copilot and Beyond~\cite{choudhuri2026copilot} & themes of AI systems developers want, and their constraints & 860 Microsoft developers; 2,580 response sets & 1--5 & GPT-5.2, Gemini 3.1 Pro, Claude Opus 4.6 & $\alpha$ per theme 0.81--0.97, mean 0.94 & Two researchers reviewed every theme and negotiated the codebook; two spot-checked rationales \\
\rowcolor{gray!10}What Makes a Great Co-worker~\cite{choudhuri2026coworker} & checking the authors' coding; coding rating reasons and work etiquette & 22 interviews; 3,334 rating reflections; 991 etiquette answers & 3--5 (researchers coded first) & Claude Opus 5, GPT-5.6, Gemini 3.6 Flash & $\alpha$ 0.84 (transcripts); median 0.88 (reasons), 0.94 (etiquette) & Two researchers reviewed every theme and negotiated the codebook; two spot-checked rationales \\
You Shall Not Pass~\cite{choudhuri2026youshallnotpass} & classifying responses onto levels of AI autonomy & 448 developers; 1,535 task-level records & 4--5 (fixed rules) & GPT-5.4, Gemini 3 Flash, Claude~Sonnet~4.6 & three-rater $\alpha$ 0.93 (two researchers and the council) & Two researchers coded 67 random response sets: human--human $\kappa$ 0.94; human--council $\kappa$ 0.95 \\
\bottomrule
\end{tabularx}
\vspace{-2mm}
\end{table*}

\textbf{\textit{Stage 5: Inter-rater reliability and consensus.}} Finally, this stage combines the three AI coders' decisions and checks whether each code was applied consistently across the corpus. First, a deterministic script settles each unit's code assignments by majority vote: a code that at least two of the three AI coders assign \emph{reaches consensus} on that unit and enters the coded corpus. Units that at least two AI coders code \textsc{issue} are excluded from prevalence counts.

Second, the script measures whether the three AI coders apply each code in the same way. It computes Krippendorff's $\alpha$ for each code~\cite{krippendorff2018content,mcdonald2019reliability} and reports the median, minimum and maximum across codes, which show how agreement varies. It also reports pairwise Cohen's $\kappa$~\cite{cohen1960coefficient}, which shows whether one AI coder diverges from the other two. In \full{}, $\alpha$ is also computed on the assignments after the adversaries' critiques. Codes below the $\alpha$ threshold go back to the researchers, with the units on which the AI coders disagreed, for the decision set out in Stage~0.

Third, because AI coders can agree on the wrong code and $\alpha$ cannot necessarily catch this~\cite{kohli2026nine}, researchers must check validity. Specifically, they code the held-out sample (Stage~0) without seeing the consensus coding, resolve their differences, and compare their coding with the consensus codes using Cohen's $\kappa$ (\emph{human--council agreement}). Finally, the researchers spot-check a random sample of rationales against the codebook, review the units coded \textsc{issue} or \textsc{uncovered} and, in \full{}, the high-uncertainty assignments, and use their discretion to decide what to do with them.

\textbf{\textit{Output: coded corpus.}} \tact{} outputs the coded corpus, e.g., as a spreadsheet, together with the approved codebook. Each unit is tagged with the codes that reached consensus, the AI coders' rationales and the representative quotes, such that each tag can be traced back to the respondent.

\textbf{\textit{\lite{} in the field.}} We distilled \tact{} from three of our own prior studies, which ran all or part of \lite{} on corpora of up to 3,300 responses (Table~\ref{tab:studies}). The first full run of \lite{} was in~\cite{choudhuri2026copilot}. In~\cite{choudhuri2026coworker}, we ran Stages~3--5 after the researchers had coded the data themselves, and in~\cite{choudhuri2026youshallnotpass}, we ran Stages~4--5 with a finalized codebook. Table~\ref{tab:studies} describes each study, so that readers can see how \lite{} was implemented in different settings and contexts. Furthermore, in this paper, we validate \tact{} on an independent dataset that other researchers collected, analyzed, and published in 2023, giving us an independent reference coding to compare against, as we detail next.

\section{Validation on an Independent Dataset}
\label{sec:demo}

In this section, we report our experiment testing whether the \tact{} procedure reproduces an independent study's coding, to establish its criterion validity. We used the dataset from Zhang et al.'s 2023 study of how developers use GitHub Copilot~\cite{zhang2023demystifying}. Their coding is our reference, and we measured the council's agreement with it using recall, precision, and F1. We chose this study because its authors coded the data manually and released both the data and their coding~\cite{zhang2023dataset}, so the reference is independent of any model, and others can replicate our comparison with the threads and mapping we release.

\vspace{-1mm}
\subsection{The reference study}

Zhang et al.\ collected Stack Overflow posts and GitHub Discussions about Copilot as of 18 June 2023~\cite{zhang2023demystifying}. On Stack Overflow, their search for ``copilot'' returned 678 unique posts. After two authors piloted the criterion that a post must give information about Copilot (Cohen's $\kappa = 0.773$), they retained \dmZhangThreads{}. From each post (question, answers and comments), they extracted eight data items, one per research question (RQ), recording an item only where a post stated it explicitly. Three RQs, specifically the programming languages, IDEs and technologies used with Copilot, were counts suited to quantitative analysis. The remaining five RQs were suited to qualitative analysis, and the authors analyzed them by constant comparison: (1)~the \textbf{functions} developers implement with Copilot, (2)~their \textbf{purposes} for using it, (3)~its \textbf{benefits}, (4)~the \textbf{limitations and challenges} they face, and (5)~their \textbf{expected features}. Two authors coded the posts, the first author reviewed the other's coding and grouped the codes into categories, and a third author examined the coding and categories. The three discussed every divergence until they reached negotiated agreement. The result is a category scheme for each of the five RQs, with every coded item linked to its source text.

\vspace{-1mm}
\subsection{Validation study design}

\textbf{\textit{Data and research questions.}} Zhang et al.\ collected data from two sources and released each as a separate file, along with their coding~\cite{zhang2023dataset}. For our validation, we scoped our analysis to one source, their Stack Overflow data, and to their five qualitative RQs, for two reasons: (1)~\textbf{coverage}: the Stack Overflow data alone carried labels for all five RQs (Table~\ref{tab:demo-results}); and (2)~\textbf{cost}: it kept the cost of running both variants stage by stage modest. Adding the GitHub Discussions would have roughly quadrupled the corpus without substantively changing what the comparison tests.

Zhang et al.\ released the IDs and links of their \dmZhangThreads{} Stack Overflow posts and their coding, as one row per coded excerpt with its category. Using the IDs, we retrieved each post's text, answers, comments and revision history through the Stack Exchange API. Of the \dmZhangThreads{} posts, \dmCorpusThreads{} still existed and \dmDeletedThreads{} had been deleted since 2023. Because Zhang et al.'s reported frequencies combine both sources, we validated against the labels they gave each post instead of their frequencies.

We treated each thread, a post with its answers and comments, as our unit of analysis. We then rebuilt each thread as it stood on 18 June 2023, replaced its source-code snippets with a placeholder, and prompted the AI analysts with the five RQs and their goals and rationales, in Zhang et al.'s original wording (see supplement~\cite{tactsupplement}).

The AI analysts read all \dmCorpusThreads{} threads during discovery. We reserved \dmDevThreads{} of them as the pool from which we drew the blind-pass sample and the codebook's examples (Section~\ref{sec:demo-run}). We computed every agreement and reliability figure on the other \dmEvalThreads{} \emph{evaluation threads}.

\textbf{\textit{Setting up \tact{}.}} We compared \lite{} and \full{} in a controlled, stage-by-stage design. At each stage, we first ran \lite{} and then added \full{}'s components to its output, so that both variants started from the same input and any difference at that stage came from those components.

We used instances of Claude Opus~5, GPT-5.6 and Gemini 3.6 Flash as the AI analysts and the AI coders, and an instance of Claude Opus~5 as the reconciler. As adversaries, instances of Gemini critiqued Claude, Claude critiqued GPT, and GPT critiqued Gemini, and an instance of GPT-5.6 critiqued the reconciler. In Stage~4, we used TypeSafe's Jev~1.13 as the decision model~\cite{typesafe2026systemone}. We ran the AI analysts and the reconciler at each provider's default reasoning mode (adaptive thinking for Claude, auto mode for Gemini and GPT). No model saw Zhang et al.'s labels.

\textbf{\textit{Comparison with the reference.}} We compared the council's codes with Zhang et al.'s \emph{categories}, i.e., the themes in their scheme (e.g., \emph{difficulty of integration} under limitations and challenges). Zhang et al.'s released coding listed each coded excerpt with its category, but not the post it came from. After approving the codebook, we therefore located each excerpt in the threads and kept only the labels whose excerpt appeared in exactly one evaluation thread. This resulted in a total of \dmEvalLabels{} \emph{reference labels}, each a category given to one thread. Because a thread can carry labels for several RQs, these reference labels mapped to \dmEvalLabelledThreads{} of the \dmEvalThreads{} evaluation threads. The remaining evaluation threads carried no reference label, primarily because Zhang et al.\ recorded no item for them on the five RQs.

The council's codes and Zhang et al.'s categories also divided the data differently, so we mapped each council code to their categories using only the definitions. A council code mapped to every category that a researcher using Zhang et al.'s scheme might reasonably have assigned to a thread carrying that code. An independent instance of Claude Opus~5 built the mapping with a rationale for each council code and reconciled it with another independent mapping by a new instance of GPT-5.6. Neither saw Zhang et al.'s labels. Two authors from our research team further spot-checked the rationales and froze the mapping before scoring began.

We scored agreement at two stages. At the codebook stage, a council code counted for every thread it cited. At the coding stage, it counted only for the threads where it reached consensus. The \dmEvalLabels{} reference labels spanned 43 of Zhang et al.'s categories, and the final codebook had a code for 42 of them, so one label was dropped, and the coding stage scored against the \dmFCLabels{} labels in those 42 categories. We considered a reference label \emph{matched} when its thread carried a council code mapped to the label's category. \emph{Recall} and \emph{precision} are, respectively, the share of reference labels matched and the share of the council's assignments that mapped to a category Zhang et al.\ gave the same thread, counted only on threads they labeled for the same RQ. F1 combines the two and measures agreement only on the codes each side applied, which suits a reference that records only what was stated explicitly~\cite{hripcsak2005agreement}. Because a mapping that sent every council code to every category would make precision 1.0, we report each mapping's breadth (categories per code) alongside precision.

\textbf{\textit{Single-model baseline.}} As a further point of comparison, we also ran what a team might do today with an off-the-shelf LLM in a chat window. We gave GPT-5.6 all \dmCorpusThreads{} threads and the five RQs and asked it to build a codebook and code every thread. We then mapped its codes and scored them in the same way (Section~\ref{sec:demo-results}).

\vspace{-1mm}
\subsection{Running \tact{}}\label{sec:demo-run}

\textbf{\textit{Data preparation.}} We coded the corpus in batches of 25 threads. A script first drew the blind-pass sample, 50 seeded random threads (the default minimum) from the \dmDevThreads{} reserved threads, and the held-out sample, 50 seeded random threads from the evaluation threads.

\textbf{\textit{Discovery and reconciliation.}} Under \lite{}, the three AI analysts made \dmCallsLiteDisc{} discovery calls (Table~\ref{tab:demo-cost}). They discovered 130, 58 and 110 codes and listed 17, 32 and 9 threads, respectively, as having no code. The provenance check traced all \dmProvQuotes{} quotes to their cited threads and confirmed that no thread was skipped. The reconciler then merged these codes into a \lite{} candidate codebook of \dmCodesLiteCandidate{} codes.

Under \full{}, the discovery adversaries raised 48, 37 and 48 critiques, mostly about weakly substantiated or miscounted support, or codes that should be merged or split. The AI analysts fully accepted 24, 34 and 39 of these critiques, partly accepted 13, 3 and 7, and rejected 13 in all; the critiques and answers added 12 calls, for \dmCallsFullDisc{} in all. The reconciliation adversary raised 26 critiques, mostly about merging codes with little overlap in their cited threads. The reconciler fully accepted 10, partly accepted 14 and rejected 2, producing a \full{} candidate codebook of \dmCodesFullCandidate{} codes.

\textbf{\textit{Human review and codebook approval.}} Before seeing any model output, two authors from our research team each open-coded the blind-pass sample. The blind pass surfaced no code missing from the council's candidate codes. We then reviewed each code in both candidate codebooks against its cited evidence, settled our differences through negotiated agreement, and recorded each edit and its reason in the review log. We dropped weakly supported codes, merged similar ones and approved a first codebook of 80 codes.

\textbf{\textit{Coding and consensus.}} The three AI coders then coded all \dmCorpusThreads{} threads with the first codebook. In this first coding round, 34 of the 80 codes fell below the $\alpha$ threshold and were returned with the threads on which the AI coders disagreed. Following our decision rule in Stage~0, we consolidated them, approved a final codebook of \dmCodesReviewed{} codes, and reran all coding. After the rerun, we decided to keep every final code, so that both variants were scored on the same codes; Table~\ref{tab:demo-results} reports their $\alpha$ as measured. Each variant coded the corpus with this final codebook. \lite{}'s coding is therefore not an end-to-end \lite{} run; we chose this design so that we could reliably attribute any difference in coding to \full{}'s coding components (screening and the coding adversaries). Accordingly, we compare the two variants in two parts: their candidate codebooks (Stages~1--3) and their coding with the same final codebook (Stages~4--5), see Table~\ref{tab:demo-results}.

On the evaluation threads under \full{}, at least two AI coders coded \dmCodeFullIssue{} threads \textsc{issue}, such as threads not about Copilot, and no \textsc{uncovered} content reached consensus. To check that the screening component in \full{} (recall that in Stage 4, the decision model screens the codes and shows each AI coder only those likely to apply to a unit) did not hide codes the AI coders would otherwise have assigned, we measured its recall: of the codes that reached consensus under \lite{}, where the AI coders saw every code, \dmScreenShareShown{}\% were among the codes that screening showed them under \full{}.

Finally, two authors from our research team spot-checked the AI coders' rationales on a random 10\% of threads and found none that disagreed with the codebook. They also reviewed each high-uncertainty assignment and coded the held-out sample, whose agreement with the council we report in Section~\ref{sec:demo-results}.

\begin{table}[t]
\footnotesize
\caption{Agreement of \lite{} and \full{} with Zhang et al.'s labels~\cite{zhang2023demystifying} at the codebook and coding stages (RQ: research question; R: recall; P: precision), and the AI coders' median $\alpha$.}
\label{tab:demo-results}
\setlength{\tabcolsep}{1.7pt}
\begin{tabularx}{\columnwidth}{@{}>{\raggedright\arraybackslash}Xlrcccrcccc@{}}
\toprule
& & \multicolumn{4}{c}{Codebook} & \multicolumn{4}{c}{Coding} & \\
\cmidrule(lr){3-6}\cmidrule(lr){7-10}
RQ & & Labels & R & P & F1 & Labels & R & P & F1 & $\alpha$ \\
\midrule
Functions & \lite{} & 27 & 0.70 & 0.59 & 0.64 & 27 & 0.59 & 0.68 & 0.63 & 0.83 \\
 & \full{} &  & 0.74 & 0.66 & 0.70 &  & 0.59 & 0.68 & 0.63 & 0.94 \\
\addlinespace[2pt]
Purposes & \lite{} & 24 & 0.88 & 0.65 & 0.75 & 24 & 0.79 & 0.82 & 0.81 & 0.77 \\
 & \full{} &  & 0.96 & 0.70 & 0.81 &  & 0.83 & 0.86 & 0.85 & 0.83 \\
\addlinespace[2pt]
Benefits & \lite{} & 21 & 0.86 & 0.77 & 0.81 & 20 & 0.70 & 0.76 & 0.73 & 0.64 \\
 & \full{} &  & 0.81 & 0.74 & 0.78 &  & 0.70 & 0.81 & 0.75 & 0.78 \\
\addlinespace[2pt]
Limitations & \lite{} & 94 & 0.82 & 0.70 & 0.75 & 94 & 0.66 & 0.68 & 0.67 & 0.89 \\
 & \full{} &  & 0.83 & 0.69 & 0.76 &  & 0.68 & 0.67 & 0.68 & 0.92 \\
\addlinespace[2pt]
Exp.\ features & \lite{} & 17 & 0.94 & 0.67 & 0.78 & 17 & 0.94 & 0.80 & 0.86 & 0.77 \\
 & \full{} &  & 0.94 & 0.81 & 0.87 &  & 0.94 & 0.77 & 0.85 & 0.94 \\
\addlinespace[2pt]
\midrule
Overall & \lite{} & 183 & 0.82 & 0.68 & 0.75 & 182 & 0.70 & 0.72 & 0.71 & 0.77 \\
 & \full{} & & \textbf{0.84} & \textbf{0.71} & \textbf{0.77} &  & \textbf{0.71} & \textbf{0.72} & \textbf{0.72} & \textbf{0.87} \\
\bottomrule
\end{tabularx}
\vspace{-2mm}
\end{table}

\vspace{-1mm}
\subsection{Results}\label{sec:demo-results}

\textbf{\textit{Agreement of the candidate codebooks.}} At the codebook stage, \lite{}'s candidate codebook matched \dmCBLiteReached{} of the \dmEvalLabels{} reference labels (recall \dmCBLiteRecall{}, precision \dmCBLitePrecision{}, F1 \dmCBLiteFone{}), and \full{}'s matched \dmCBFullReached{} (recall \dmCBFullRecall{}, precision \dmCBFullPrecision{}, F1 \dmCBFullFone{}; Table~\ref{tab:demo-results}). Because a broader mapping makes a match easier, we also report each mapping's breadth, the average number of Zhang et al.'s categories to which a council code was mapped; a lower breadth means a stricter mapping. \lite{}'s mapping averaged \dmMapLiteCatsPerCode{} categories per code and \full{}'s \dmMapFullCatsPerCode{}, so \full{}'s higher scores came even under the stricter mapping.

In both variants, precision was lower than recall, partly because the council's codes were finer-grained than Zhang et al.'s categories~\cite{zhang2023demystifying}. A council code counted against precision when it cited a thread to which Zhang et al.\ gave a neighboring category. Under \full{}, 98 code--thread pairs counted against precision in this way, 54 of them among the limitation categories. There, the council divided single categories, most often \emph{limits of code generation} or \emph{difficulty of integration}, into distinct problems, such as Copilot stopping without an error, Copilot overriding editor settings and key bindings, disabling Copilot as the only remedy, and corporate proxies blocking access. We list all 98 pairs in the supplement for readers to judge; some reflected differences in granularity or threads that raised several points. Another plausible cause for the finer distinctions is cognitive load. Research on working memory has shown that people can hold only a few items in mind at once~\cite{miller1956magical,cowan2001magical,sweller1988cognitive}, so applying dozens of categories across hundreds of long threads can strain a human coder, while a coarser scheme keeps that load manageable. Because the council sees every code's definition in every call, it may separate accounts that a human team would otherwise merge into one category.

\textbf{\textit{Agreement of the final coding.}} At the coding stage, \lite{}'s final coding matched \dmFCLiteReached{} of the \dmFCLabels{} reference labels (recall \dmFCLiteRecall{}, precision \dmFCLitePrecision{}, F1 \dmFCLiteFone{}), and \full{}'s matched \dmFCFullReached{} (recall \dmFCFullRecall{}, precision \dmFCFullPrecision{}, F1 \dmFCFullFone{}; Table~\ref{tab:demo-results}), against \dmBaseMatched{} for the single-model baseline (recall \dmBaseRecall{}, precision \dmBasePrecision{}, F1 \dmBaseFone{}). Both variants performed better than the baseline: in 1,000 paired bootstrap resamples of the evaluation threads, \lite{} and \full{} exceeded the baseline's F1 in \dmBaseBootLite{}\% and \dmBaseBootFull{}\% of resamples, respectively. Overall recall at the coding stage was lower than at the codebook stage because a council code counted only on threads where it reached consensus, whereas the codebook stage credited every thread an AI analyst cited. Of Zhang et al.'s \dmZhangCategories{} categories, 46 had a corresponding council code. Conversely, \dmMapRevUnmapped{} of the \dmCodesReviewed{} final codes, such as the expectation that Copilot work behind a company proxy, had no category in their scheme, so none of their assignments could match a reference label. Table~\ref{tab:demo-results} also reports agreement for each of Zhang et al.'s five RQs.

\textbf{\textit{Coding reliability.}} The median per-code Krippendorff's $\alpha$ among the three AI coders was \dmCodeLiteAlphaMedian{} under \lite{} and \dmCodeFullAlphaMedian{} under \full{}, both above the default threshold of 0.7 (Section~\ref{sec:procedure}). The median per-code Cohen's $\kappa$, averaged over the three pairs of AI coders, was \dmCodeLiteKappa{} under \lite{} and \dmCodeFullKappa{} under \full{}. No single AI coder diverged substantively from the other two: under \lite{}, the median pairwise $\kappa$ was \dmCodeLiteKappaClGp{} for Claude--GPT, \dmCodeLiteKappaClGe{} for Claude--Gemini and \dmCodeLiteKappaGpGe{} for GPT--Gemini (under \full{}, \dmCodeFullKappaClGp{}, \dmCodeFullKappaClGe{} and \dmCodeFullKappaGpGe{}). Median $\alpha$ was higher under \full{} on every RQ (Table~\ref{tab:demo-results}).

\textbf{\textit{Human--council agreement.}} On the 50 held-out threads, which two authors from our research team coded independently with the final codebook, without seeing the model assignments, the council’s coding agreed with ours at a Cohen’s $\kappa$ of 0.79 under \lite{} and 0.85 under \full{}, pooled over all (thread, code) decisions on the \dmCodesReviewed{} final codes. Whereas Zhang et al.’s coding tested the council against another team’s, this sample tested it against our own.

\begin{table}[h]
\small
\caption{Model calls, input and output tokens (millions) and cost (USD) of \lite{} and \full{} by stage for the 256 threads, at list prices in September 2026. Calls and tokens count generative model calls; \full{}'s costs also include the decision model's screening requests.}
\label{tab:demo-cost}
\setlength{\tabcolsep}{2pt}\footnotesize
\begin{tabularx}{\columnwidth}{@{}Yrrrrrrrr@{}}
\toprule
& \multicolumn{4}{c}{\lite{}} & \multicolumn{4}{c}{\full{}} \\
\cmidrule(lr){2-5}\cmidrule(l){6-9}
Stage & Calls & In & Out & \$ & Calls & In & Out & \$ \\
\midrule
Discovery & 19 & 3.26 & 0.64 & 23.02 & 31 & 6.79 & 1.11 & 48.69 \\
Reconciliation & 5 & 0.30 & 0.15 & 5.14 & 15 & 1.12 & 0.33 & 12.64 \\
Coding & 33 & 1.59 & 0.07 & 7.10 & 99 & 4.95 & 0.34 & 24.67 \\
\midrule
Total & 57 & 5.15 & 0.85 & 35.26 & 145 & 12.86 & 1.78 & 86.00 \\
\bottomrule
\end{tabularx}
\vspace{-2mm}
\end{table}

\textbf{\textit{Token economics.}} The council ran on all \dmCorpusThreads{} threads, although we scored agreement on the \dmEvalThreads{} evaluation threads. Model calls cost \$\dmCostLiteTotal{} under \lite{} and \$\dmCostFullTotal{} under \full{} (Table~\ref{tab:demo-cost}), or \$\dmCostLitePerHundred{} and \$\dmCostFullPerHundred{} per 100 threads. These per-100 figures apply to corpora of similar length, because coding cost scales with the number of threads and discovery cost with the length of the corpus and the number of RQs. The decision model's calls, one request per thread scoring all \dmCodesReviewed{} final codes, cost \$\dmCostJevScreen{}.

\section{Discussion}
\label{sec:discussion}

Our study suggests that \tact{} can carry out the labor-intensive part of qualitative analysis that often scales with the number of responses, and can thus help SE researchers and practitioners scale such analysis while retaining final authority over interpretation. In this section, we discuss three common questions that we have received about this process: (A)~What can go wrong, and how can a team guard against it? (B)~How much does it cost, and can a team afford it? (C)~How can a team use \tact{} within its own practice?

\vspace{-1mm}
\subsection{What can go wrong, and how can a team guard against it?}
Based on our experience developing \tact{}, Table~\ref{tab:pitfalls} lists common pitfalls in AI-assisted qualitative analysis across stages, why each matters, and an example safeguard that we built into \tact{} against it. Three of them warrant discussion because they can undermine an otherwise reliable analysis. %

\textbf{\textit{(1) Evidence that looks right.}} Consider a quote that a model has edited into a paraphrase, or a code that cites units that do not support it. Agreement among the AI coders measures only whether they assign the same codes, so high agreement among models may not surface either problem. Toward this end, \tact{} checks every cited unit and quote against the corpus using deterministic scripts, and researchers then judge whether each quote supports its code.

\textbf{\textit{(2) Anchoring effect.}} Researchers who read the candidate codes before the data can come to see the data through those codes, a form of anchoring bias~\cite{schroeder2025just}. Their agreement with \tact{} may then reflect what they were shown, not an independent reading of what respondents said. We therefore advocate that researchers form their own reading first, as in \tact{}'s blind pass, and only then review the candidate codes.

\textbf{\textit{(3) Shared model error.}} AI coders that see each other's output, or that share a model family, tend to agree more than independent models would~\cite{kim2025correlated,kohli2026nine}, and their consensus cannot always expose an error they share~\cite{liu2026agreement}. \tact{} therefore draws its AI coders from three distinct model families and keeps them independent. Because independent models can still err together, in \tact{}, we also recommend that researchers check the council against their own coding of a held-out sample.

\begin{table*}[t]
\footnotesize
\caption{Common pitfalls in AI-assisted qualitative analysis, why they matter, and example safeguards, each built into \tact{}.}
\label{tab:pitfalls}
\setlength{\tabcolsep}{4pt}\renewcommand{\arraystretch}{1.0}
\begin{tabularx}{\textwidth}{@{}>{\raggedright\arraybackslash}p{1.75cm}>{\raggedright\arraybackslash}p{3.7cm}>{\raggedright\arraybackslash}X>{\raggedright\arraybackslash}X@{}}
\toprule
Stage & Common pitfall & Why it matters & Example safeguard \\
\midrule
\textbf{Discovery} & Accepting evidence that is almost right & Edited quotes and inflated support lists pass every agreement check, because AI coders agree on them as readily as on exact ones & Check every identifier and quote against the corpus by program; researchers then judge whether the quotes support the code \\
\rowcolor{gray!10} & Trusting that a model analyzed the whole corpus & A reported count of units analyzed can be correct while the analysis was partial & Require every unit to be cited by a code or labeled as having no code; analyze large corpora in batches \\
 & Ignoring order, prompt and batch effects & Code lists depend on the prompt and on the order of units~\cite{wang2024notfair} & Give every AI analyst one versioned prompt and the same three analysis layers \\
\midrule
\rowcolor{gray!10}\textbf{Reconciliation} & Merging on similar wording, or averaging opposed positions & Disagreement between AI analysts disappears before reliability is computed & Merge only codes that share units and describe the same thing, with a rationale for each group in the merge plan, which researchers review \\
\midrule
\textbf{Human review} & Reviewing the model codes before coding the data independently & Patterns no model proposed stay unseen, and model codes anchor researchers~\cite{schroeder2025just,gao2024coaicoder} & Researchers code a sample before seeing model output, note what the council lacks, and settle codes by negotiated agreement \\
\midrule
\rowcolor{gray!10}\textbf{Coding} & Revising the codebook during coding & Early and late batches follow different schemes & Close the codebook before coding, and code new content \textsc{uncovered} for the researchers \\
 & Silently dropping off-topic or ambiguous responses & A response one AI coder drops, and another codes, is a disagreement that never reaches the statistics & Code such responses \textsc{issue}, exclude them from prevalence counts only by consensus, and report the count \\
\rowcolor{gray!10} & Letting AI coders see each other, or drawing them from one model family & Imitation and shared training inflate agreement~\cite{kim2025correlated} & Code independently, one AI coder from each of three model families, and report per-code $\alpha$ \\
\midrule
\textbf{Reliability} & Reading model agreement as correctness & Consensus cannot expose an error that all AI coders share~\cite{liu2026agreement} & Researchers code a held-out sample without model output; report human--council agreement separately \\
\midrule
\rowcolor{gray!10}\textbf{Adversaries} & Letting adversaries edit the output they critique & The adversary becomes a fourth AI analyst, and independence is lost & Adversaries only critique; the model whose output was critiqued decides \\
\bottomrule
\end{tabularx}
\end{table*}

\subsection{How much does it cost, and can a team afford it?}
Using state-of-the-art LLMs available at the time, our validation cost \$\dmCostLiteTotal{} in model calls under \lite{} and \$\dmCostFullTotal{} under \full{} for \dmCorpusThreads{} threads (Section~\ref{sec:demo-results}). These costs bear on a recurring concern that AI in research will widen the gap between researchers who can pay for the most capable models and those who cannot. One counterargument is that the price of a given level of model performance has fallen 9- to 900-fold per year, depending on the task~\cite{epoch2025llmprices}. We do not take this trend as settling the concern. Prices can rise, access to the most capable models differs across countries and institutions, and a cost that is modest for one team can be prohibitive for another. We therefore built \tact{} so that it does not depend on any single model provider or on its most expensive configuration.

Toward this end, we recommend that teams start from \lite{} with three model families, which we call the reference configuration. In our validation, \full{} gave the more rigorous checks, raising the AI coders' agreement and bringing its candidate codebook closer to the reference coding, while \lite{}'s final coding agreed with the reference coding nearly as closely, at less than half the cost. Teams with the resources can add \full{}'s components, and we recommend adding the discovery adversaries first, since a code that the AI analysts miss is harder to recover later than a coding error, which the other two AI coders expose. Where budgets, company policy or the need to keep data on the team's own servers rule out commercial models from three providers, a team could draw the three AI coders from three open-weight model families run locally and check them against a held-out sample in the same way. Finally, whether to use \tact{} at all should rest on the size of the corpus, the number of research questions and the team's time. For a corpus that two researchers can code within a week, \tact{} may add little, whereas for larger corpora, it can let a team analyze data it would otherwise set aside.

\subsection{How can a team use \tact{} within its own practice?}
\tact{} need not be used whole. We built it for thematic analysis~\cite{braun2006using}, and a thematic analysis can use every stage, as ours did. Teams that follow other approaches may still find individual stages useful. For example, in the open coding phase of grounded theory~\cite{glaser1967discovery,charmaz2014constructing}, a team might use discovery and reconciliation to surface candidate codes across a large corpus, while memoing, theoretical sampling and theory building stay with the researchers. A team that already has a coding frame, as in deductive coding, might use only the coding stages, as we did in an earlier study that classified responses onto predefined levels~\cite{choudhuri2026youshallnotpass}. To support such uses, we release our pipeline, with its prompts and programs for every stage~\cite{tactsupplement}.

Whatever parts a team adopts, we recommend that it preserve provenance, so that someone outside the analysis can trace how each finding was reached~\cite{lincoln1985naturalistic}. \tact{} can enable human analysts to trace any finding back to the respondents' own words, to see which model proposed its code and on what evidence, and to see how the codes were modified and why. While provenance cannot show that an interpretation is right, it lets others examine how the interpretation was reached, which is what trustworthiness in qualitative research fundamentally asks for~\cite{lincoln1985naturalistic}.

Future work should investigate how \tact{} performs in other languages and with open-weight models. Furthermore, the community should ask how reliance on models for parts of qualitative analysis affects human analysts, for example whether they still know what the data says and still empathize with respondents as they would after carefully reading every account.

\vspace{-1mm}
\subsection{Threats to validity}
\label{sec:limitations}

As with any study, our validation had threats to validity. Our use of AI may have introduced codes that reflect what the models tend to produce more than what respondents meant~\cite{bano2024large,martinezmontes2025large}. Our mitigations were three of \tact{}'s safeguards: AI analysts from three model families proposed codes independently, a deterministic script checked every quote against the corpus, and we reviewed every candidate code against its cited evidence.

Because our validation used a public dataset, the models may have seen it during training. Zhang et al.'s threads and labels have been public since 2023, so we cannot rule out such exposure. Our main mitigation against this threat was to keep Zhang et al.'s labels from every model during the run. Also, the single-model baseline, which had the same exposure, agreed less with Zhang et al.'s coding than either variant of \tact{}, so exposure alone does not explain \tact{}'s advantage over a single model.

Our own judgments may also have shaped the results. Model suggestions can anchor the researchers who review them~\cite{schroeder2025just,gao2024coaicoder}, and the two authors from our research team who coded the held-out sample had designed \tact{}. Our mitigations were to complete the blind pass before seeing any model output, to record every refinement to the codebook along with its rationale in the review log, and to code the held-out sample without seeing the council's coding. Zhang et al.'s coding gave a second check that did not depend on our judgment. Only our mapping of the council's codes to their categories did, and we froze that mapping before the evaluation.

Finally, our earlier studies drew on industry survey and interview data, and our validation drew on one English-language forum, so we do not claim that our results hold universally across corpora, languages or models. Instead, we offer \tact{} as a starting point for teams wanting to conduct such analysis rigorously at scale. To support replication, we have made the \tact{} pipeline publicly available~\cite{tactsupplement}.

\section{Conclusion}

\tact{} lets SE researchers and practitioners analyze qualitative responses reliably and economically at scale, while retaining final authority over interpretation. On an independent dataset that other researchers had coded manually, \tact{}'s candidate codebook agreed with the reference coding at an F1 of \dmCBLiteFone{}(Lite)--\dmCBFullFone{}(Full), and its final coding of the corpus at an F1 of \dmFCLiteFone{}--\dmFCFullFone{}, while performing better than a single-model baseline given the same task. Furthermore, \tact{}'s AI coders agreed with one another at a median Krippendorff's $\alpha$ of \dmCodeLiteAlphaMedian{}--\dmCodeFullAlphaMedian{}, and their coding agreed with our own coding of a held-out sample at Cohen's $\kappa$ of 0.79--0.85, indicating consistent and reliable coding. \tact{} was also economical. Model calls cost \$\dmCostLitePerHundred{}--\$\dmCostFullPerHundred{} per 100 threads (about 23,000--57,000 tokens per thread).

We invite teams to start from \lite{}, adopt its stages wholly or in part, and preserve the provenance that lets anyone trace each finding back to respondents' accounts. In this division of labor, AI models carry the scale of the analysis and researchers keep its interpretive meaning, \emph{so that teams can hear from every stakeholder who took the time to respond, and do so tactfully.}

\textbf{Data availability.} The \tact{} pipeline, alongside its prompts, validation corpus, validation and reliability scripts and dashboards, is available at~\cite{tactsupplement}.

\bibliographystyle{IEEEtran}
{\footnotesize\bibliography{references}}

\end{document}